\documentstyle[preprint,pre,aps,12pt]{revtex}
\begin{document}
\title{ Motion of Curves and Surfaces and Nonlinear Evolution Equations in 
(2+1) Dimensions }
\author{M. Lakshmanan$^{a,*}$, R. Myrzakulov$^{b,c,\dag}$, S. Vijayalakshmi$^a$ and
A.K. Danlybaeva$^c$}
\address{
$^a$Centre for Nonlinear Dynamics, Department of Physics, Bharathidasan University, 
Tiruchirapalli 620 024, India \\
$^b$High Energy Physics Institute, National Academy of Sciences, 
480082,  Alma-Ata-82, Kazakstan \\
$^c$Centre for Nonlinear Problems, PO Box 30, 480035, Alma-Ata-35, Kazakstan}
\maketitle 
\begin{abstract}
It is shown that a class of important integrable nonlinear evolution equations
in (2+1) dimensions can be associated with the motion of space curves endowed
with an extra spatial variable or equivalently, moving surfaces. Geometrical 
invariants then define topological conserved quantities. Underlying evolution 
equations are shown to be associated with a triad of linear equations. Our 
examples include Ishimori equation and Myrzakulov equations which are shown 
to be geometrically equivalent to Davey-Stewartson and Zakharov -Strachan
(2+1) dimensional nonlinear Schr\"odinger equations respectively. 
\end{abstract}
\vskip 5pt
\hskip 50pt{PACS Numbers: 02.30.Jr, 52.35.Sb, 75.10.H, 02.40.Hw}
\pacs{}

The motion of curves in $E^3$ and the defining equations of surfaces have
drawn wide interest in the past, especially since they give nice geometric 
interpretations of nonlinear evolution equations in (1+1) dimensions [1-6].
There are many physical situations where they play a natural role. Particularly
they have important connections with soliton equations solvable by Ablowitz, 
Kaup, Newell and Segur (AKNS) formalism [1-6]. For a recent renaissance of 
interest, see for example [7] and References therein. Other examples include 
dynamics of interfaces, surfaces and fronts, vortex filaments, supercoiled DNAs, 
magnetic fluxes, deformation of membranes, dynamics of proteins, propagation 
of flame fronts, and so on [8,9]. Of special interest among these systems is 
the dynamics of isotropic Heisenberg ferromagnetic spin chains where an 
interesting equivalence with nonlinear Schr\"odinger family of equations arise 
in a natural and physical way [10].

The question then arises as to whether nonlinear evolution equations in (2+1)
dimensions of importance, especially spin equations, can be given a similar 
geometrical setting associated with motion of curves endowed with an additional 
coordinate or equivalently motion of surfaces. In this letter, we develop a 
general theory of obtaining nonlinear evolution equations in (2+1) dimensions
as the compatibility requirements of the geometric equations defining the 
motion of curves and surfaces, incidentlly equivalent to a triad of linear 
equations. Our analysis also brings out certain natural topological integral 
invariants associated with the resultant systems. We then consider a class of 
(2+1) dimensional spin systems including (2+1) dimensional generalized 
Heisenberg ferromagnatic spin systems and deduce their equivalent nonlinear 
Schr\"odinger family including Davey-Stewartson, Zakharov-Strachan and other 
equations. In this context, we also wish to add that in recent times a general
approach (mainly due to Konopelchenko) has appeared in the literature[11, 12]
concerning the interpretation of hierarchies of integrable (2+1) dimensional
systems as special motions of surfaces in $E^3$ by inducing the surfaces in it
via the solutions to two dimensional linear problems (2D LPs). But in the 
present paper, we are interested in extending the theory of moving space curve formalism in (1+1) dimensions to
(2+1) dimensions.
    
We consider the space curve in $E^3$ as in Fig. 1, defined by the Serret-Frenet 
equations [13]
\begin{equation}
{\vec e}_{jx} = {\vec D} \wedge {\vec e}_{j},\,\,\,\, {\vec D} = \tau {\vec e}_{1}+ 
              \kappa {\vec e}_3,\,\,\,\, j=1,2,3 \label{eq1}
\end{equation}
where ${\vec e}_{1}, {\vec e}_{2}$ and ${\vec e}_{3}$ are the unit tangent, normal 
and binormal vectors respectively to the curve. Here $x$ is the arclength 
parametrising the curve. The curvature and torsion of the curve are defined 
respectively as
\begin{eqnarray}
\kappa & = & ({\vec e}_{1x}\cdot {\vec e}_{1x})^{1\over 2},\nonumber \\
\tau   & = & \kappa^{-2} {\vec e}_{1}\cdot ({\vec e}_{1x} \wedge {\vec e}_{1xx}) .
             \label{eq2}
\end{eqnarray}
The unit tangent vector  ${\vec e}_{1}$ is given by ${\vec e}_{1} = \frac{\partial 
{\vec r}}{\partial x}=\frac{1}{\sqrt g}\frac{\partial {\vec r}}{\partial \theta}$
where $g$ is the metric $g=\frac{\partial {\vec r}}{\partial \theta} 
\frac{\partial {\vec r}}{\partial \theta}$ on the curve such that $x(\theta,t) = 
\int\limits_{0}^{\theta}\sqrt{g(\theta^{\prime}, t)}d\theta^{\prime}$. Here 
$\theta $ defines a smooth curve  and ${\vec r}(\theta, t)$ is the position 
vector of a point on the curve at time $t$. Considering the motion of such a 
twisted curve, the time evolution of the orthogonal trihedral can be easily 
seen [3] to be 
\begin{eqnarray}
{\vec e}_{jt} & = & {\vec \Omega} \wedge {\vec e}_{j},\nonumber \\             
{\vec \Omega} & = & \omega_{1}{\vec e}_{1} + \omega_{2}{\vec e}_{2}       
                + \omega_{3}{\vec e}_{3},\,\,\,\,j=1,2,3  \label{eq3}
\end{eqnarray}
where $\omega_i, i=1,2,3$ are some functions of $\kappa $ and $\tau $ and 
their derivatives. Then the compatibility of Eqs. (1) and (3) 
gives the evolution equations for  $\kappa $  and $\tau $ [3] as
\begin{eqnarray}
\kappa_t    & = & \omega_{3x} + \tau \omega_2, \nonumber \\ 
\tau_t      & = & \omega_{1x} - \kappa \omega_2, \nonumber \\ 
\omega_{2x} & = & \tau \omega_3-\kappa \omega_1 . \label{eq4}
\end{eqnarray}
   
Now let us pass on to the subject of this letter namely to the theory of curves 
in (2+1) dimensions by endowing them with an additional spatial variable $y$.
Alternately this could represent a surface specified by the trihedral ${\vec e}_
j(x,y,t)$ which is set in motion. A starting point of our approach is the  
observation that the $y$-evolution of the trihedral ${\vec e}_{j}$ is determined by the following system of 
linear equations [14] (see also Refs. [15,16]);
\begin{eqnarray} 
{\vec e}_{jy}    & = & {\vec {\Gamma}} \wedge {\vec e}_{j}, \nonumber \\
{\vec {\Gamma }} & = & \gamma_{1}{\vec e}_{1} + \gamma_{2}{\vec e}_{2}       
                     + \gamma_{3}{\vec e}_{3} .  \label{eq5}
\end{eqnarray}
Here $\gamma_{j}$'s  are functions to be determined. Using the fact that 
${\vec e}_i$'s, $i=1,2,3$ form an orthogonal trihedral and 
using (1) and (5) in the compatibility condition ${\vec e}_{jxy}
={\vec e}_{jyx}$, we obtain
\begin{mathletters} 
\label{eq6all}
\begin{equation} 
\gamma_1  =  \partial_x^{-1}(\tau_y + \kappa\gamma_{2}),\,\,\,\,
\gamma_3  =  \partial_x^{-1}(\kappa_y - \tau \gamma_{2}),\label{eq6a} 
\end{equation} 
and $\gamma_{2}(x,y,t)$ is a solution of the following equation
\begin{equation} 
-\kappa \gamma_2  =  \tau_{y} - \left[ \frac{\tau\partial^{-1}_{x}(\kappa_y 
                       - \tau \gamma_{2})-\gamma_{2x}}{\kappa}\right]_{x},
                       \label{eq6b} 
\end{equation} 
\end{mathletters}
where $\partial_x^{-1} \equiv \int_{-\infty }^x \,dx$. We also note that Eqs.
(1) and (5) may be identified with the Gauss-Weingarten equations of surface theory
written in terms of orthogonal coordinates and so Eqs. (6) with the 
Codazzi-Mainardi equations. On the other hand, the condition ${\vec e}_{jty} = 
{\vec e}_{jyt} $ gives rise to the set of equations
\begin{eqnarray} 
\gamma_{1t} & = & \omega_{1y}+\omega_{3}\gamma_{2} - \omega_{2}\gamma_{3},
                  \nonumber \\
\gamma_{2t} & = & \omega_{2y}+\omega_{1}\gamma_{3} - \omega_{3}\gamma_{1},
                  \nonumber \\
\gamma_{3t} & = &\omega_{3y} + \omega_{2}\gamma_{1} - \omega_{1}\gamma_{2}. 
                  \label{eq7}  
\end{eqnarray}

Choosing $\gamma_1, \gamma_2, \gamma_3$ consistently so that Eqs. (4) 
and (7) are compatible, one can obtain (2+1) dimensional evolution 
equations for $\kappa$ and $\tau $. Further one can easily show that the 
defining equations for the motion of the trihedral, namely (1), 
(3) and (5) are equivalent to a set of three Riccati equations
\begin{mathletters} 
\label{eqno8all}
\begin{eqnarray} 
z_{lx}  =  -i\tau z_l+{\kappa \over 2}\left[ 1+z_l^2\right],\,\,\,\,\,\,\,\,\,\,\ 
\,\,\,\,\,\,\,\,\,\,\,\,\,\,\,\,\,\,\,\,\,\,\,\,\,\,\,\,\,\,\,\,\,\,\,\,\,\label{eq8a}\\  
z_{ly}  =  -i\gamma_1z_l+{1\over 2}\left[ \gamma_3+i\gamma_2\right] z_l^2
           +{1\over2}\left[ \gamma_3-i\gamma_2\right], \label{eq8b}
\end{eqnarray}   
and 
\begin{equation}
z_{lt}  =  -i\omega_1z_l+{1\over 2}\left[ \omega_3+i\omega_2\right] z_l^2+
            {1\over2}\left[ \omega_3-i\omega_2\right],\label{eq8c}
\end{equation}   
\end{mathletters}
where $z_l$, $l=1,2,3 $ is a scalar variable obtained through an orthogonal
rotation of the trihedral, $z_l= (e_{2l}+ie_{3l})/(1-e_{1l})$, $e_{1l}^2+
e_{2l}^2+e_{3l}^2=1$. Introducing the transformation $z_l=v_2 / v_1$, 
we obtain the triad of equivalent linear equations:
\begin{mathletters} 
\label{eq9all}
\begin{eqnarray}
v_{1x} & = & {i\tau \over 2}v_1-{\kappa \over 2}v_2,\nonumber \\ 
v_{2x} & = & {\kappa \over 2}v_1-{i\tau \over 2}v_2,\label{eq9a}\\ 
v_{1y} & = & {i\gamma_1\over 2}v_1-{1\over 2} \left( \gamma_3
             +i\gamma_2\right)v_2,\nonumber \\
v_{2y} & = & {1\over 2}\left( \gamma_3-i\gamma_2\right)v_1-{i\gamma_1\over 2} 
             v_2, \label{eq9b}\\
v_{1t} & = & {i\omega_1\over 2}v_1-{1\over 2}\left( \omega_3+i\omega_2\right) 
             v_2,\nonumber \\ 
v_{2t} & = & {1\over 2}\left( \omega_3-i\omega_2\right)v_1-{i\omega_1\over 2} 
           v_2. \label{eq9c}
\end{eqnarray}   
\end{mathletters}
The compatibility of these linear equations again gives rise to Eqs. (4), (6) 
and (7). Thus any nonlinear evolution equation obtained through the space curve
formulation in (2+1) dimensions is equivalent to a triad of linear equations.
Specific examples are given below.

The above formulation then leads us in a natural way to certain topological 
invariants as given in the following theorem:\\
{\bf Theorem}. The above constructed evolution equations in (2+1) dimensions
possess the following integrals of motion [14]
\begin{mathletters} 
\label{eq10all}
\begin{eqnarray}
K_{1}  =  \frac{1}{4\pi}\int\int \kappa \, \gamma_{2}dxdy,\nonumber \\     
K_{2}  =  \frac{1}{4\pi}\int\int \tau \, \gamma_{2}dxdy.\label{eq10a}
\end{eqnarray}

Proof: The proof is straightforward and follows from the various relations 
discussed above. For example, from (6a) we get $(-\kappa \,\gamma_2)_t
=(\tau_t)_y-(\gamma_{1t})_x$, or $(-\kappa \,\gamma_2)_t=(\omega_{1x} - \kappa 
\omega_2)_y - (\omega_{1y} + \omega_3\gamma_2 - \omega_2\gamma_3)_x$. Hence follows the 
first statement of the theorem. Similarly the second one may also be proved. 
In terms of the unit vector ${\vec e}_1$ these integrals of motion take the form
\[
K_{1} = \frac{1}{4\pi}\int\int \,\, {\vec e}_1\cdot ({\vec e}_{1x} \wedge {\vec e}
        _{1y}) dxdy,\]
and
\begin{equation}
K_{2} = 
       \frac{1}{4\pi}\int \int \frac{[{\vec e}_{1}\cdot ({\vec e}_{1x} \wedge 
       {\vec e}_{1y})][{\vec e}_{1}\cdot({\vec e}_{1x}\wedge {\vec e}_{1xx})]}
              (\vec e_{1x}^2)^{\frac{3}{2}} dxdy .\label{eq10b}
\end{equation}
\end{mathletters} 

So the above quantities $K_{j}$ are the conserved integrals - invariants 
of the (2+1) dimensional evolution equations and they play an important role 
in the theory. In particular, curves can be classified by the value of the 
topological invariants $K_{j}$ but we will not pursue this aspect further. 
We now present a few applications of the above presented formalism in finding 
the geometrically equivalent counterparts of some known (2+1) dimensional spin 
systems. To this end, firstly, following the Ref. [10] we identify 
the tangent vector ${\vec e}_1$ with the unit spin vector  ${\vec e}_1 \equiv 
{\vec S}(x,y,t)$. Secondly, we introduce the following complex transformation 
which is a generalization of the one given in Ref. [10],
\begin{equation}
q(x,y,t)  = a(x,y,t) \exp{ib(x,y,t)}, \label{eq11}
\end{equation}
where $a(x,y,t)$ and $b(x,y,t)$ are functions of $\kappa $ and $\tau $, to be 
determined, and obtain the equation for the complex function $q(x,y,t)$ which
turns out to be an interesting integrable (2+1) dimensional nonlinear evolution 
equation. We want to demonstrate our approach in the following examples of the  (2+1) 
dimensional spin systems.\\                      
{\bf A) The Myrzakulov I (M-I) equation} 

This equation reads as [14,15]
\begin{mathletters}
\label{eq12all}
\begin{eqnarray}
{\vec e}_{1t} & = & ({\vec e}_1 \wedge {\vec e}_{1y} + u {\vec e}_1)_x,\,\,\,\,
{\vec e}_1     =  {\vec S}(x,y,t),\label{eq12a}\\
    u_x      & = & - {\vec e}_1 \cdot ({\vec e}_{1x} \wedge {\vec e}_{1y}).\label{eq12b}
\end{eqnarray}
\end{mathletters} 
In this case, we obtain

\begin{mathletters} 
\label{eq13all}
\begin{eqnarray} 
\gamma_2    & = &  \frac{u_{x}}{\kappa},  \label{eq13a}\\
{\vec \Omega} & = & (\omega_{1},\,\omega_{2},\,\omega_{3}), \nonumber \\
            & = & (\frac{\kappa_{xy}}{\kappa} - \tau \partial _{x}^{-1}
                  \tau _ y ,\, -\kappa_ y ,\,- \kappa\partial^{-1}_ x \tau_ y ).
                 \label{eq13b} 
\end{eqnarray}
\end{mathletters} 
For the functions $a(x,y,t)$ and $b(x,y,t)$ in the transformation (11),
we take the form
\begin{equation}
a=\frac{\kappa(x,y,t)}{2},\,\,\, b=-\partial^{-1}_x\tau (x,y,t).\label{eq14}
\end{equation}
Then the function $q$ satisfies the following evolution equation,
\begin{equation}
iq_t (x,y,t) = q_{xy} + Vq,\,\,\,\,V_x = 2 \,(\mid q\mid^2)_y, \label{eq15}
\end{equation}
an equation which belongs to the class of equations originally discovered by 
Calogero [17] and then discussed by Zakharov [18] and 
recently rederived by Strachan [19]. Many of its properties have 
received considerable attention recently [20].\\
{\bf B) The Myrzakulov III (M-III) equation}                        

This equation which is a generalization of Eq. (12) has the form [14]
\begin{mathletters}
\label{eq16all}
\begin{eqnarray}
{\vec e}_{1t} = ({\vec e}_{1} \wedge {\vec e}_{1y} + u {\vec e}_{1})_{x} 
              + 2f(cf + d){\vec e}_{1y} - 4 c v {\vec e}_{1x},\label{eq16a}\\
u _{x} = - {\vec e}_{1} \cdot ({\vec e}_{1x} \wedge {\vec e}_{1y}),\,\,\,\, 
v _{x} = \frac{1}{4(2fc + d)^2} ( {\vec e}_{1x}^{2})_{y}, \label{eq16b}
\end{eqnarray}
\end{mathletters}
where $f$, $c$ and $d$ are constants. In this case, using the same form for 
$\gamma_i$'s and $\omega_i$'s as in Eq. (13) and with the following 
choice for the functions $a(x,y,t)$ and $b(x,y,t)$,
\begin{equation}
a(x,y,t) = \frac{\kappa(x,y,t)}{2(2cf + d)},\,\,\, b(x,y,t) = 2f(cf + d)x 
           -\partial^{-1}_{x}\tau ,\label{eq17}
\end{equation}
we obtain the evolution equation
\begin{equation}
iq_t (x,y,t) = q_{xy} - 4ic (Vq)_x + 2d^{2} Vq,\,\,\,\,   
               V_x = (\mid q\mid^2)_y   \label{eq18}
\end{equation} 
This equation as well as the M-III equation (16) are integrable, that 
is they have the Lax representations [14] and some properties of both 
equations were investigated in [14-16,21,22]. Note 
that Eqs. (16) and (18) admit integrable reductions: when 
$c=0$  they pass on to the M-I and Zakharov equations Eqs. (12) and 
(15) respectively. If we consider the case $d = 0$, then they reduce to 
the M-II and Strachan equations respectively [23].\\
{\bf C) The Ishimori equation}

This equation has the form [24]
\begin{mathletters}
\label{eq19all}
\begin{eqnarray}
{\vec e}_{1t}  =  {\vec e}_1\wedge ({\vec e}_{1xx} + \sigma^2{\vec e}_{yy})+
                u_x{\vec e}_{1y}+u_y{\vec e}_{1x}, \label{eq19a} \\
u_{xx}-\sigma^2 u_{yy}  =  -2\sigma^2 {\vec e}_1\cdot ({\vec e}_{1x}\wedge 
                        {\vec e}_{1y}),   \label{eq19b}
\end{eqnarray}
\end{mathletters}
where $\sigma^{2} = \pm 1$. For the case $\sigma^{2} =-1$, we obtain
\begin{mathletters}
\label{eq20all}
\begin{eqnarray} 
\gamma_2   & = & \frac{u_{xx}+u_{yy}}{-2 \kappa},
                 \nonumber \\
\omega_{1} & = & \frac{\tau \omega_3 - \omega_{2x}}{\kappa}, 
                 \label{eq20a}\\
\omega_{2} & = & u_{x} \gamma_{2} - \kappa_{x}+\gamma_{3y} + 
                 \gamma_{1}\gamma_{2}, \label{eq20b}\\
\omega_{3} & = & - \kappa \tau + \kappa u_{y} + u_{x}\gamma_{3}- 
                 \gamma_{2y} + \gamma_1\gamma_3.  \label{eq20c}
\end{eqnarray}
\end{mathletters}
If we choose $a(x,y,t)$ and $b(x,y,t)$ in the form
\begin{mathletters}
\label{eq21all}
\begin{eqnarray}
a & = & {1\over 2}[\kappa^2+\gamma^2_2+\gamma^2_3-2\kappa \gamma_{2} 
]^{1\over 2},\label{eq21a}\\
b & = & \partial_x^{-1}\left ( \tau-{u_y\over 2}+{\gamma_2\gamma_{3x}-\gamma_3
\gamma_{2x}-\kappa \gamma_{3x}\over {\kappa^2+\gamma_2^2+\gamma_3^2-2\kappa
\gamma_2}}\right),\label{eq21b}
\end{eqnarray} 
\end{mathletters}
then the function $q$ satisfies the following equation
\begin{mathletters}
\label{eq22all}
\begin{eqnarray}
iq_t (x,y,t) + q_{xx} - q_{yy} - 2q\phi = 0,\label{eq22a}\\
\phi_{xx}+\phi_{yy} + (\mid q\mid^2)_{xx}-(\mid q \mid^2)_{yy} = 0,  
\label{eq22b}
\end{eqnarray}     
\end{mathletters}
where $\phi $ is a function of $x$, $y$ and $t$, which is nothing but the 
Davey-Stewartson equation II [6]. Note that Eqs. (19) and 
(22) are known to be gauge equivalent to each other [25]. Here 
their geometrical equivalence has been estabilished. Similarly equivalence 
can be estabilished for $\sigma^2=1$.\\
{\bf D) The (2+1) dimensional isotropic Heisenberg ferromagnet model}

This equation has the form [26]
\begin{equation}
{\vec e}_{1t} = {\vec e}_{1} \wedge ({\vec e}_{1xx} + {\vec e}_{1yy}).  \label{eq23}
\end{equation}
In this case, we get
\begin{eqnarray}
\omega_{1} & = & \frac{\tau\omega_{3}-\omega_{2x}}{\kappa},\,\,\,\,
\omega_{2} = - \kappa_{x} - \gamma_{3y} - \gamma_{1}\gamma_{2} ,\nonumber \\
\omega_{3} & = & \gamma_{2y} - \kappa \tau - \gamma_{1}\gamma_{3}.\label{eq24}
\end{eqnarray}
However one finds that $\gamma_i$'s can not be uniquely found due to lack 
of an equation of the form (12b) or (16b) or (19b). So Eq. (6b) does not 
appear to be solvable, indicating that no straightforward linearization is 
available for the (2+1) dimensional isotropic Heisenberg spin equation without
introducing constraints.

In conclusion, we have shown that a geometrical formulation can be developed 
for a class of (2+1) dimensional nonlinear evolution equations through moving 
space curve formalism. The formulation shows how a class of evolution equations 
can be associated with a triad of linear equations. Whether such a connection 
alone is sufficient to prove the integrability of the underlying (2+1) dimensional
systems is a further intricate question. Even in (1+1) dimensions, there are 
some equations which though linearizable are not integrable; for example, the 
spherically symmetric Heisenberg ferromagnetic spin chain equation is 
equivalent to a nonlocal nonlinear Schr\"odinger equation which is of non-
Painlev\'e type and so nonintegrable [27]. So also is the (1+1)
dimensional isotropic Landau-Lifshitz equation with Gilbert damping [28] 
and so on. However the three examples considered in this paper are integrable 
namely the M-I, M-III and Ishimori equations and their geometrical equivalents 
namely the Zakharov, Strachan and Davey-Stewartson equations respectively are 
also integrable. Their linearization can be performed through the set of 
equations (9) and the associated linear eigenvalue problems can be constructed.
On the other hand the pure isotropic Heisenberg spin chain equation in (2+1) 
dimensions (example D above) seems not to be even linearizable by the space 
curve formalism without constraints, indicating the neceesity of additional 
scalar fields or 
nonlocal terms for its linearization. It is hoped that other (2+1) dimensional 
equations of interest like the (2+1) dimensional Korteweg-de Vries, Nizhnik- 
Novikov-Veselov, breaking soliton etc. equations and so on may have similar  
geometrical interpretations in terms of moving space curves or surfaces, for
example with restrictions on the value of the curvature $\kappa $ or torsion
$\tau $. These possibilities are being explored at present.

Finally it is of interest to consider the connection between our approach and
to that of Konopelchenko who uses the (2+1) dimensional nonlinear PDEs to
induce integrable dynamics (deformations) of the induced surfaces[11]. In our
approach, we consider the compatibility of three linear problems namely 
equations (1), (3) and (5). Two of them (eqs. (1) and (5)) can be considered
as equivalent to general surfaces in orthogonal coordinates. In the approach 
of Konopelchenko, for a given 2D LP, the variable coordinates $X^i$ ($i = 1,2,3$)
of a surface in $E^3$ are defined as some integrals over certain bilinear
combinations of solutions $\psi$ of the 2D LP and solutions $\psi^*$ of the
adjoint 2D LP and thus surfaces are induced via the solutions of 2D LPs. Then
the compatibilty of 2D LPs with time evolution leads to hierarchies of (2+1)
dimensional nonlinear evolution equations both for the coefficients of the 
2D LP and for the wave function $\psi$. This nonlinear evolution equation
induces the corresponding evolution of the induced surface. In this sense, we
believe that the two approaches differ from each other and give complementary
approaches to integrable (2+1) dimensional evolution equations.

This work of M.L. forms part of a Department of Science and Technology, 
Government of India sponsored research project. R.M. wishes to thank 
Bharathidasan University for hospitality during his visits to Tiruchirapalli. 
S.V. acknowledges the receipt of the award Junior Research Fellowship from the
Council of Scientific and Industrial Research, India.

\section*{Figure Captions}
Fig. 1 : Motion of a space curve in $E^3$ with the orthogonal trihedral
$\vec e_1$, $\vec e_2$ and $\vec e_3$
\newpage
{\bf Running Title}: Motion of curves/surfaces and (2+1) NLEEs
\pagestyle{empty}
\end{document}